\documentclass[aps,prl,twocolumn,superscriptaddress,groupedaddress,showpacs]{revtex4}
\usepackage{graphicx}  % needed for figures
\usepackage{dcolumn}   % needed for some tables
\usepackage{bm}        % for math
\usepackage{amssymb}   % for math

\newcommand{\be}{\begin{equation}}
\newcommand{\ee}{\end{equation}}
\newcommand{\bea}{\begin{eqnarray}}
\newcommand{\eea}{\end{eqnarray}}

\usepackage{setspace}
\usepackage{amsmath}
\usepackage{amssymb}
\usepackage{fancybox}
\usepackage{listings}
\usepackage{url}

\usepackage{float}
\begin{document}

\title{Condensation of Photons coupled to a Dicke Field in an Optical Microcavity}
\author{Eran Sela}
\affiliation{Raymond and Beverly Sackler School of Physics and Astronomy, Tel-Aviv University, Tel Aviv, 69978, Israel}
\author{Achim Rosch}
\affiliation{Institute for Theoretical Physics, University of Cologne, 50937 Cologne, Germany}
\author{Victor Fleurov}
\affiliation{Raymond and Beverly Sackler School of Physics and Astronomy, Tel-Aviv University, Tel Aviv, 69978, Israel}

\begin{abstract}
Motivated by recent experiments reporting Bose-Einstein condensation (BEC) of light coupled to incoherent dye molecules in a microcavity, we show that due to a dimensionality mismatch between the 2D cavity-photons and the 3D arrangement of molecules, the relevant molecular degrees of freedom are collective Dicke states rather than individual excitations. For sufficiently high dye concentration  the coupling of the  Dicke states with light will dominate over local decoherence.
This system also shows Mott criticality despite the absence of an underlying lattice in the limit when all dye molecules become excited.
\end{abstract}


\maketitle

In 1954 Dicke pointed out the possibility of collective spontaneous emission, known as superradiance~\cite{Dicke54}, in situations when $N$ nearby atoms are coupled by a single mode of the electromagnetic field. They form a collective  ``superatom'', known as the Dicke state, and emit coherently with an intensity that scales as $N^2$ instead of $N$.
A finite temperature phase transition into this superradiant phase was rigorously proven~\cite{Hepp} within the Dicke model, as was recently observed in a BEC of rubidium atoms in an optical cavity~\cite{esslinger}.

Recent experiments on an optical cavity filled with dye molecules reported room temperature BEC of photons~\cite{Klaers}. Here, two cavity mirrors at distance $d = n_z \lambda/2$ confine the light in one direction, where $n_z$ is an integer, leading to a cutoff frequency $\omega_{c} = 2 \pi c/\lambda$, where $c$ is the speed of light, above which photons disperse as two-dimensional (2D) massive particles with mass $m=\hbar \omega_{c}/c^2$~\cite{Carusotto},
 which is so small that high transition temperatures could be reached  at low photon densities.
Most importantly, the dye molecules provided a thermal reservoir equilibrating the photons on time scales short compared to any loss times of photons. At large loss and pumping rates, however,  equilibration is not efficient and standard lasing behavior takes over~\cite{Kirton13,Fischer}.

\begin{figure}[ht]
\centering
\includegraphics*[width=\columnwidth]{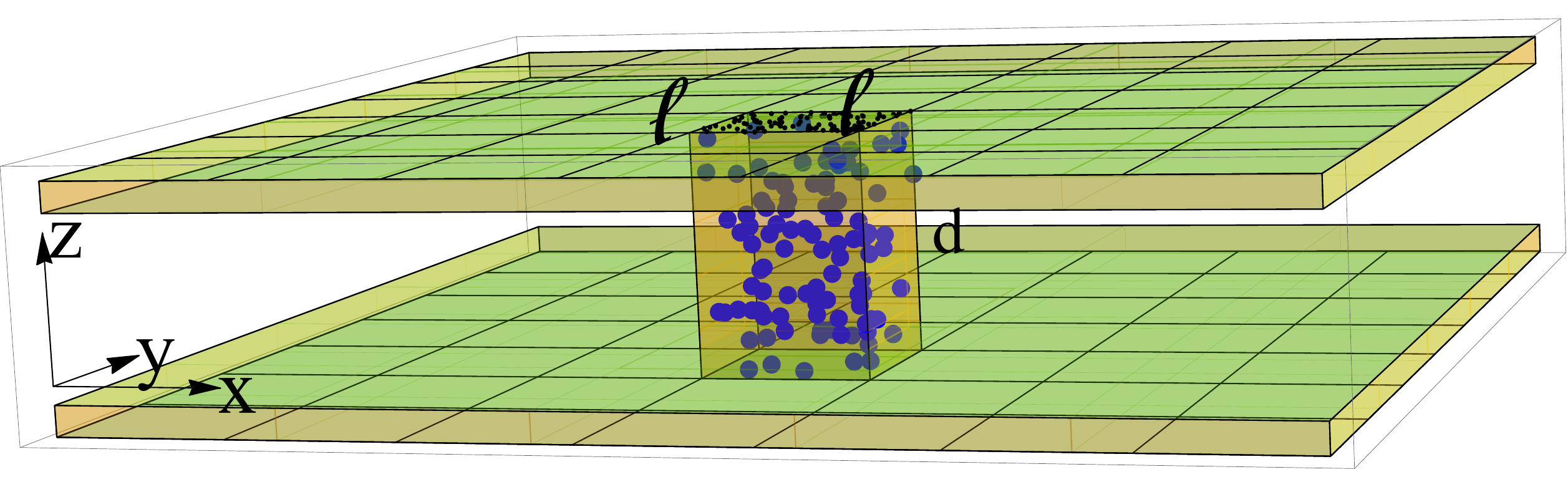}
%\vspace{-1cm}
\caption{\label{fg:2.5} Schematic view of the system and coarse graining procedure. Each box contains $N = \ell^2 d \rho_{\rm TLS}$ TLSs.
}
\end{figure}

Motivated by this experimental setup, we study the properties of 2D massive photons coupled to two-level systems (TLS).
%We  study regimes where the  coupling is almost resonant and/or where a large fraction of TLSs is excited.
As in the experiment, we assume that the typical distance $\delta_M$ between neighboring TLSs (dye molecules) is much shorter than $d$. We argue that  in close analogy to the Dicke effect, the photons interact {\em  locally} with a large number $N$ of TLSs, see Fig.~\ref{fg:2.5}. Thus locally a ``Dicke field'' forms (rather than a single Dicke state considered, e.g., in Ref.~\cite{Hepp,esslinger}), which condenses at a superfluid (SF) phase transition.

A remarkable consequence of the collective Dicke behavior is its persistence against local decoherence. In the dye molecules decoherence stems from the coupling of the TLS excitation to local ro-vibrations, and prevents coherent light-matter coupling~\cite{Martini}. However, while for $N$ molecules the (free) energy reduction due to coupling to the local decoherence degrees of freedom scales as $\propto N$, at large $N$ this coupling becomes subdominant compared to the ground state energy of the Dicke state $\propto N^2$.
\vspace{-3mm}
%Thus for high concentrations of TLSs decoherence becomes a negligible perturbation.

Our results suggest that Dicke physics is crucial in this type of systems. In the currently available experimental system in Ref.~\cite{Klaers}, the decoherence rate $\Gamma \sim 10^{14} {\rm{Hz}}$ reaches the order of the cutoff frequency $\omega_{c} = 2 \pi \times 5.1 \times 10^{14} {\rm{Hz}}$, which makes application of our model problematic. Therefore reaching  the condition $\Gamma < \omega_{c}$ may be necessary for the applicability of our theory, which then predicts a new SF regime where a BEC of ``Dicke supermolecules'' forms and facilitates overcoming decoherence.

In addition to superfluidity, the TLS filled microcavity offers a playground for Mott-Hubbard phases and quantum phase transitions (QPTs). Such phases of light were proposed to occur in arrays of photonic crystal cavities~\cite{Greentree06}, while here no artificially designed lattices are required.

\paragraph{Model:} Following Ref.~\cite{Klaers}, we formulate a model where 2D photonic excitations above the cutoff frequency, created by $f^\dagger(r)$ $[r=(x,y)]$, are converted back and forth into electronic TLS  excitations, created on the $i-$th dye molecule by Pauli matrix $\sigma^+_i$. As loss and pumping rates are assumed to be small compared to intrinsic equilibration rates, we can ignore them and consider a model with a conserved number of excitations, $\int   f^\dagger(r)  f(r) d^2 r + \sum_i (\sigma_i^z+1)/2$, controlled by the chemical potential $\mu$; see Refs. \cite{Dimer07,DallaTorre13,Buchhold13} for a discussion of the role of losses.
%, including in the Dicke model
We also consider decoherence due to a coupling to vibrational modes, created by $v^\dagger_{l,i}$. The total Hamiltonian is
\bea
\label{eq:Hmodel}
H &=&H_{\rm{ph}}+ H_{\rm TLS}+H_{\rm{TLS-ph}}+H_{\rm{vib}}, \\
H_{\rm ph}&=&\int d^2 r  f^\dagger(r)  \left( -\frac{{\hbar^2 \mathbf \nabla}_\textbf{r}^2}{2 m}  -\mu \right)f(r),\nonumber \\
H_{{\rm TLS}}&=&-(\Delta+\mu) \sum_{i}  (\sigma^z_i+1)/2,\nonumber \\
H_{\rm TLS-ph} &=&\sum_{i}(g_{i} f^\dagger(r_i) \sigma^-_{i}+h.c.),\nonumber \\
H_{\rm{vib}} &=& \sum_i [ \sum_l (E_l v_{l,i}^\dagger v_{l,i}+\sigma^z_i (C_l  v_{l,i}+h.c.))].\nonumber
\eea
The molecules are located at random 3D positions, $ R_i =(r_i,z_i)$, with density $\rho_{\rm TLS}=\delta_M^{-3}$, and $g_i \propto \sin(n_z \pi z_i/d)$.
Eq.~(\ref{eq:Hmodel}), based on the rotating wave approximation, assumes that the detuning $\Delta$ is much smaller than $\hbar \omega_{c}$.

To compare to the experimental setup of  Ref.~\cite{Klaers} we relate the  parameters of Eq.~(\ref{eq:Hmodel})  to measurable quantities. Consider an excited state with $\sigma^z_i = 1$ for $\Delta<0$. A transition to the ground state, $\sigma^z_i = -1$, through photon emission occurs according to Fermi's golden rule with the rate  $\frac{1}{\tau}  = \frac{m g^2}{\hbar^3}$. A second measurable rate concerns the decay of off-diagonal elements of the density matrix of a TLS~\cite{Martini}. This decoherence rate is related to the transition between the superposition states $\sigma^x_i = \pm 1$ occurring due to the coupling to the local vibrations, and within our model at $T=0$ this rate from excited- to groundstate is given by  $\Gamma =\frac{2 \pi}{\hbar} \sum_l C_l^2 \delta(E_l- \Delta E)$, where $\Delta E$ is the relevant energy difference between $\sigma^x_i = \pm 1$.
Quoting  typical numbers from Ref.~\cite{Klaers,Klaers1},
\be
\label{eq:numbers}
\delta_M = 10\,{ nm}, d=1.5\, \mu m,n_z=7,  \tau \sim 1\, ns, \Gamma \sim 10^{14}{\rm{Hz}},
\ee
we observe that the decoherence rate is much larger than the photon-matter coupling $\Gamma \tau \gg 1$ and that $d/\delta_M \gg 1$. In the following, we will first study the problem for $\Gamma=0$, ignoring $H_{\rm{vib}}$. For $d/\delta_M \gg 1$ we show that collective Dicke states control the phase diagram. In a second step, we investigate under which conditions the collective Dicke excitations remain coherent even for $\Gamma \tau \gg 1$.

\paragraph{Emergence of Dicke states:} We now derive the effective 2D theory describing the dispersive 2D photons coupled to the static randomly located TLSs in the limit $\delta_M \ll d$.
%This theory will be defined for a  coarse graining scale $\ell$.
%, on which both the photon field $f(r)$ and the average two level system state $\langle \sigma_i \rangle$ vary slowly, which is much larger than the 2D distance between impurities.
%It is natural to associate the length scale $\ell$ with an effective spatial extent to individual TLS excitations in the $xy$ plane due to their coupling with the photon field.
%We consider large detuning $\Delta <0$.
%To see this,
%To construct this theory,
We first consider a limit with large $\Delta>0$ and $\Delta+\mu\lesssim 0$ where only a few TLSs are excited (but no photons) such that $\sigma^z_i \cong -1$
for almost all TLSs. In this case one can replace the Pauli operators $\sigma^+_i$ and $\sigma^-_i$ in Eq.~(\ref{eq:Hmodel}), satisfying $[\sigma^+_i , \sigma^-_j]=\delta_{ij} \sigma^z_i \cong -\delta_{ij}$, by conventional bosonic creation and annihilation operators, $a^\dagger_i$ and $a_i$, using $(\sigma^z_i+1)/2=a^\dagger_i a_i$
In this limit the effective single-particle spectrum can be determined from~\cite{Haldane76,Fleurov76}
$\det [(E+\Delta) \delta_{ij} -g_i I_{ij}(E)  g_j]=0$, with
\be
\label{eq:hopping}
I_{ij}(E) = \int \frac{d^2 k}{(2 \pi)^2} \frac{e^{i k (r_i - r_j)}}{E-\hbar^2 k^2/(2 m)}.
\ee
The photon propagator $g_i I_{ij} g_j$  describes hopping of excitations between TLSs via virtual photons. Here $I_{ij}$ decays on the length scale
 \bea
\label{eq:ell}
\ell = \hbar/\sqrt{ 2 m |\Delta|},
\eea
with $I_{ij}(E) \approx I_{ij}(-\Delta) \propto e^{-|r_i-r_j|/\ell}$.
When the 2D distance parallel to the mirrors is large,  $|r_i - r_j| \gg \ell$, their direct coupling can be neglected.
In contrast, for $|r_i - r_j| \ll \ell$, one obtains $I_{ij}(E) \approx I(-\Delta)$ independent of $i$ and $j$. In this case
the effective hopping Hamiltonian $\sum_{ij} a_i^\dagger g_i I g_j a_j= g_{\rm eff}^2 I \alpha^\dagger \alpha$ describes the coupling of a single collective state $\alpha^\dagger =\sum_i g_i a_i^\dagger /\sqrt{\sum_i g_i^2} $ with enhanced coupling strength
$g_{\rm eff}=\sqrt{\sum_i g_i^2} = \sqrt{N} (\langle g^2 \rangle)^{1/2}$ where $N$ is the number of states with $|r_i - r_j| \lesssim \ell$~\cite{Delta0}.

Therefore each region of typical linear dimension $\ell$ in the $xy$ plane, containing $N=d {\ell}^2 \rho_{\rm TLS}$ TLSs, gives $N-1$ dark states that are essentially decoupled from the continuum, and a single bright state, referred to as the Dicke state, gains kinetic energy and propagates along the microcavity. The latter couples to the continuum with an effective coupling enlarged by $\sqrt{N}$,
\be
\label{eq:lambdaeff}
g_{\rm{eff}} =c' \sqrt{N} \sqrt{ \langle g^2 \rangle},
\ee
where $c'$ is a numerical coefficient of order $1$.

\begin{figure}[h]
\begin{center}
\includegraphics*[width=\columnwidth]{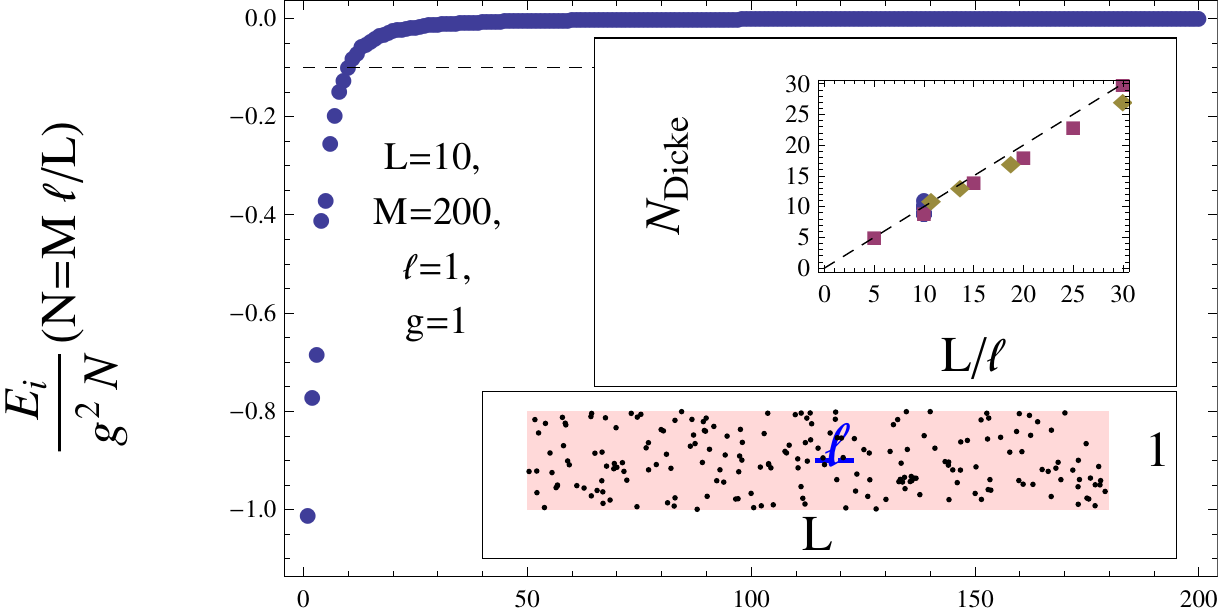}
%\vspace{-1cm}
\caption{\label{fg:energiesANDsystem} Eigenenergies of $H'$ (in increasing order) for a disorder realization. The plot in the inset shows, for a variety of values of $M$, $L$ and $\ell$, that the number of hopping particles, defined as the set of energies satisfying $E_i \le -0.1 g^2 N$ from the main plot, scales as $N_{Dicke} \propto L/\ell$.
}
\end{center}
\end{figure}

We have checked that this picture holds independent of disorder realization, for a technically simpler but conceptually identical problem, of 1D photons coupled to 2D molecules. We diagonalized numerically the disordered single particle Hamiltonian $H' = \sum_{ij} a^\dagger_i g_i I_{ij} g_j a_j$ where the hopping amplitude $I_{ij} = e^{-|x_i-x_j|/\ell}$ depends only on the $x$ coordinate, $g_i = g \sin (\pi n_z y_i)$, and a set of $M$ points $(x_i,y_i)$ is randomly located on a two dimensional strip of length $L$ and width $1$, see lower inset of Fig.~\ref{fg:energiesANDsystem}. As shown in Fig.~\ref{fg:energiesANDsystem},
the energy eigenstates $E_i$ can be roughly divided into two types, according to
\bea
E_i \sim
\begin{cases}
 0 & \text{if }i \in \text{dark~states}, \\
  \mathcal{O}(g_{\rm{eff}}^2) & \text{if }i \in \text{Dicke~states}.
\end{cases}
\eea
Therefore all dark states are localized within length $\ell$ or shorter, and the $\sim L/\ell$ Dicke states form a band of extended states with bandwidth $\sim g_{\rm{eff}}^2$.

\paragraph{Phase diagram:} We will now construct the zero temperature phase diagram, as a function of the detuning $\Delta$ and chemical potential $\mu$. As long as $\mu$ and $\mu+\Delta$ are sufficiently negative, we expect a fully gapped state without excitations (vacuum state). Upon increasing $\Delta$ or $\mu$ the energy gap will eventually close at a QPT. Having identified the Dicke states as the lowest energy excitations in the regime with $\langle \sigma^z_i \rangle \approx 1$, we now formulate a coarse-grained effective theory for long wavelength $\gg \ell$ that will allow us to determine the transition lines where the Dicke excitations become gapless. Keeping only the Dicke state $\alpha_{\cal B} = \frac{\sum_{i \in {\cal B}} g_i a_i}{\sqrt{\sum_{i \in {\cal B}} g_i^2}}$ for each box ${\cal B}$ of area $\ell^2$ (see Fig.~\ref{fg:2.5}) and letting $f(r_i)|_{i \in {\cal B}} \approx f(r_{\cal B})$ where $r_{\cal B}$ is the center of box ${\cal B}$, we obtain
$H_{\rm TLS}^{\rm{eff}}=- (\Delta + \mu) \sum_{\cal B}  \alpha^\dagger_{\cal B} \alpha_{\cal B}$ and $H_{\rm{TLS-ph}}^{\rm{eff}} =\sum_{\cal B} g_{\rm{eff},\mathcal{B}} f^\dagger(r_{\cal B}) \alpha_{\cal B} +h.c.$, where $g_{\rm{eff},{\cal B}} =\sqrt{\sum_{i \in {\cal B}}g_i^2}$. We now turn to a continuum theory,  by introducing a \emph{Dicke field} $D(r) \equiv \alpha_{\cal B}(r_{\cal B})/\ell$, such that $\int d^2 r D^\dagger D =\sum_{\cal B} \alpha_{\cal B}^\dagger \alpha_{\cal B}$. The effective Lagrangian density
%, related to the Hamiltonian density by $\mathcal{L} = i \hbar \partial_t- \mathcal{H}$,
becomes
\bea
\label{eq:lagrangian22}
\mathcal{L}_{0} =\Phi^\dagger \left(
\begin{array}{cc}
i \hbar \partial_t+  \frac{\hbar^2}{ 2 m} \nabla_r^2+\mu & -\Omega \\
-\Omega & i \hbar \partial_t+\mu+\Delta \\
\end{array}
\right) \Phi,
\eea
where $\Phi=(f(r),D(r))^T$. The collective energy scale
 \be
\label{eq:Omegadef} \Omega=g_{\rm{eff}} \ell^{-1} =c' \sqrt{\langle g^2 \rangle \rho_{\rm TLS} d},
  \ee
appears, which does not depend on the coarse-graining length $\ell$ ---  only the 2D projected density of TLSs, $\rho_{TLS}d$, enters. As desired, the dispersion of excitations can be extracted from Eq.~(\ref{eq:lagrangian22}), setting $\det \mathcal{L}_{0}[\omega (k), k]=0$; we find that the gap vanishes along the line $\mu_{0} = -\frac{\Delta}{2} - \sqrt{\left(\frac{\Delta}{2} \right)^2+{\Omega}^2}$ (see
 %the phase diagram in
 Fig.~\ref{fg:1}).
\begin{figure}[h]
\centering
\includegraphics*[width=6cm, angle=270]{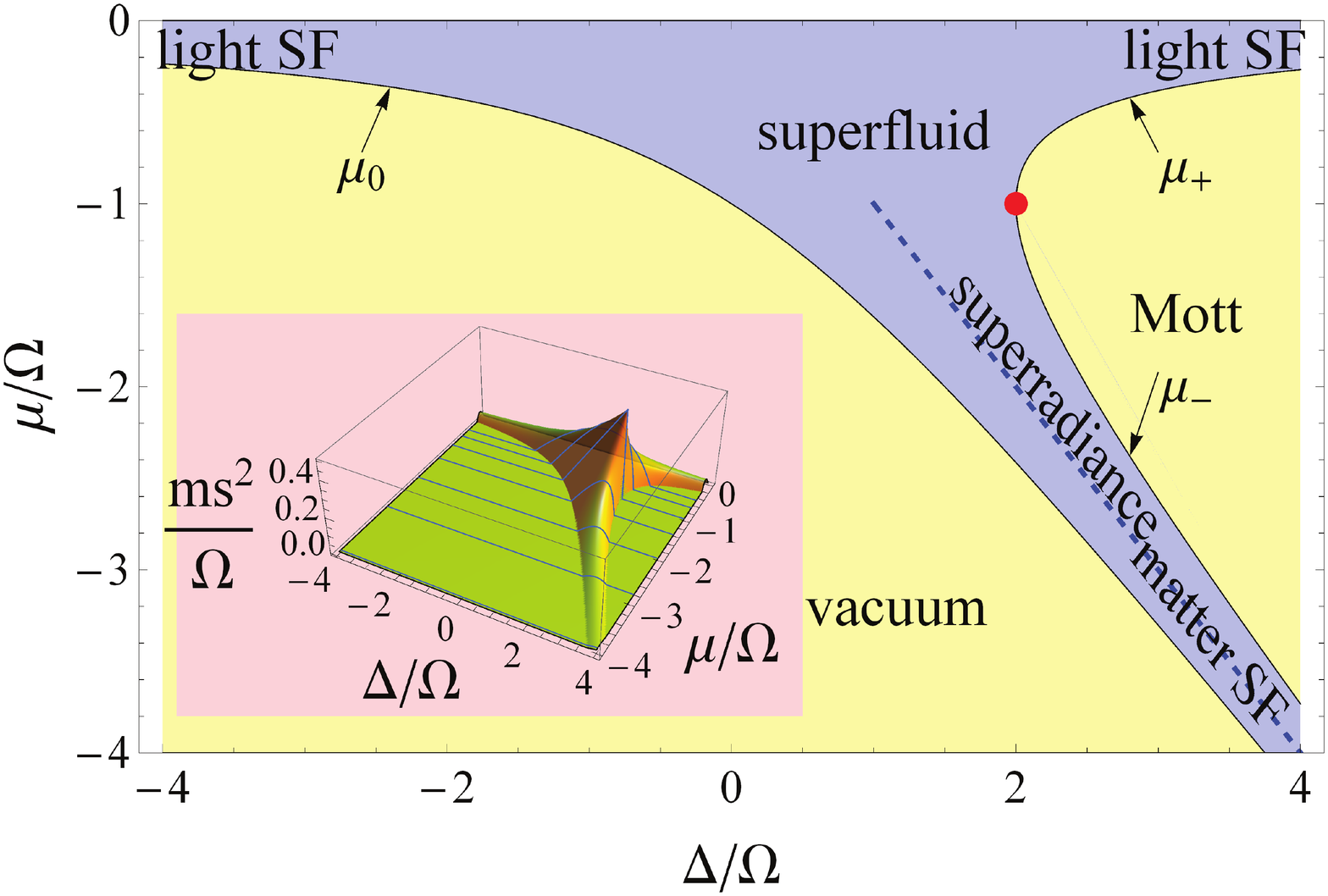}
%\vspace{-1cm}
\caption{\label{fg:1} Phase diagram of model (\ref{eq:Hmodel}) at $T=\Gamma=0$.
The Mott phase describes a state where all TLSs are excited, the red point marks the Lorentz invariant  point characteristic of a Mott transition.
Condensation is strongly enhanced by the Dicke effect in a regime where approximately half of the TLSs are excited (dashed line).
  Inset: SF velocity given by Eq.~(\ref{eq:sgeneral}).
}
\end{figure}

\vspace{-1mm}
Similarly,  consider a state where all TLSs, but no other photon modes, are excited. In this state the number of excitations coincides with the number of TLSs. It can be identified with a Mott phase at filling $1$. Its range of stability is easily obtained similarly to the vacuum state. Now we are in the opposite limit $\langle \sigma^z_i \rangle\cong +1$, and here we may define a Dicke state $\bar{D}$ corresponding to a linear superposition of deexcited TLSs. The resulting Lagrangian is
\bea
\label{eq:Lprime}
\mathcal{L}_{1} =\bar{\Phi}^\dagger \left(
                \begin{array}{cc}
                  i \hbar \partial_t+  \frac{\hbar^2}{ 2 m} \nabla_r^2+\mu & -\Omega \\
                  -\Omega & -i \hbar \partial_t-\mu-\Delta \\
                \end{array}
              \right) \bar{\Phi},
              \eea
where $\bar{\Phi}=(f(r),\bar{D}^*(r))^T$. From this equation we find that the Mott phase is only stable for $\Delta > 2 \Omega$ and $\mu_- < \mu < \mu_+$ with  $\mu_{\pm} = \frac{1}{2} \left(-\Delta \pm \sqrt{\Delta^2 - 4 {\Omega}^2} \right)$, see
 Fig.~\ref{fg:1}.

\vspace{-1mm}
At $T=0$ one obtains  superfluidity  (Fig.~\ref{fg:1}) between the vacuum and the Mott phase.
The critical properties of the QPT from the vacuum or the Mott phase into the superfluid state are determined by the
quadratic dispersion of the first boson, which condenses,\emph{ i.e.}, $\omega(k)= (\mu_c-\mu)+ \hbar^2 k^2 /(2 m^*) +\mathcal{O}(k^4)$~\cite{appendix}. As $\omega \sim k^2$ at criticality, the dynamical critical exponent is  $z=2$.
%(and interactions are marginally irrelevant in $d+z=4$).
A characteristic feature of the Mott phase~\cite{Sachdev} is that there is always a point where the effective mass changes its sign and therefore vanishes. Here this occurs at $\Delta= 2 \Omega$, $\mu=-\Omega$ (point in Fig.~\ref{fg:1}), where Eq.~(\ref{eq:Lprime}) gives rise to a Lorentz invariant critical mode with linear dispersion ($z=1$), $\omega(k)=c_{\rm eff} k$ with
\bea
\label{stip}
c_{\rm eff} =\left( \frac{\Omega}{2m}\right)^{1/2}.
\eea
$c_{\rm eff}$ can become very large [$c_{\rm eff}\approx 0.1 c$ using the parameters of Eq.~(\ref{eq:numbers})] but remains always smaller than $c$ \cite{appendix}.
In this case the universality class of the QPT is given by the 3D XY model. Close to this point, one can also expect the emergence of a Higgs mode \cite{Higgs}.

\paragraph{Superfluid regime and large spin physics:}

To describe the ground state of the condensed Dicke states and photons, which can no longer be described as a dilute gas, we map the Dicke states onto large spins, approximating~\cite{remark} the local coupling constants $g_i$ by a constant $g$. We define a  large-$S$ operator for each correlated region ${\cal B}$, $\vec{S}_{\cal B} =\sum_{i \in {\cal B}} \frac{1}{2} \vec{\sigma}_{i}$. Then the TLS Hamiltonian becomes
\bea
\label{dyeph}
 H_{\rm TLS}^{\rm{eff}}+H_{\rm{TLS-ph}}^{\rm{eff}}&=& -\sum_{\cal B}  (\mu+\Delta)  S_{\cal B}^z  \\
 &+& \sum_{\cal B}  g [ f^\dagger(r_{\cal B}) S^-_{\cal B} +S^+_{\cal B} f(r_{\cal B})]. \nonumber
\eea
We will apply the standard mean field approximation in order to decouple the spin-photon interaction, relegating details to the Supplement~\cite{appendix}.
%In fact this mean field approximation becomes exact at large-$S$.
As in the Dicke model~\cite{Hepp}, inside the SF phase in Fig.~\ref{fg:1}, we obtain a condensed state where the experimentally measurable photon order parameter $\langle f^\dagger \rangle $ is linked to the spin-projection into the $xy$ plane, $\sqrt{{S^x}^2+{S^y}^2} =  |\mu \langle f^\dagger  \rangle| \ell^2 /g$, and given by
\be
\label{eq:f2}
|\langle  f^\dagger \rangle|^2 =\frac{1}{4}\left( \frac{\Omega^2}{\mu^2}-\frac{(\mu+\Delta)^2}{\Omega^2} \right) \rho_{\rm TLS} d.
\ee
Deep in the condensed phase, we observe that along the line $\Delta+\mu=0$ (dashed line in Fig.~\ref{fg:1}) we have $\langle S^z \rangle=0$, and the intensity of condensed photons scales with the square of the number of dye molecules, $|\langle f^\dagger \rangle |^2 \propto (\rho_{\rm TLS} d)^2$, similar to Dicke superradiance. Also the groundstate energy density, given by
\be
\label{eq:EGS}
E_{GS} =\frac{ \mu^2 (\Delta+\mu)^2+\Omega^4 }{4 \mu \Omega^2} \rho_{TLS}d,~~(\mu <0)
\ee
scales as $-(\rho_{TLS} d)^2$ for $\Delta+\mu=0$.

In contrast to the Dicke model studied in \cite{Hepp} with all TLSs located at a single point, in our model slow fluctuations around the homogenous solution result in a linearly dispersing Goldstone mode with velocity
\be
\label{eq:sgeneral}
s = \sqrt{\frac{\mu^3 (\Delta+\mu)^2 - \mu \Omega^4}{2m(2 \Delta \mu^3+3 \mu^4+\Omega^4)}} \xrightarrow[\Delta = 2 \Omega, \mu=-\Omega]{} c_{\rm{eff}},
\ee
plotted in the inset of Fig.~\ref{fg:1}, and reproducing Eq.~(\ref{stip}) at the Lorentz invariant critical point.
%, %$\Delta/\Omega=2$, $\mu/\Omega=-1$,
%however similar velocities can be obtained in a broad range of parameters.
This finite SF velocity marks the role of interactions, and pinpoints the principal difference between the local Dicke model~\cite{Hepp} and our interacting BEC of a Dicke field.

\vspace{-1mm}
The Hilbert space of a box with $N$ TLSs ($2^N$ states) is much greater than the $2S+1=N+1$ states of a spin-$S$ object. The additional small spin states with $\vec{S}_{\cal B}^2 < S (S+1)$, have higher energy similar to Fig.~\ref{fg:energiesANDsystem}. We integrate out the high energy photon modes from Eq.~(\ref{dyeph}), giving~\cite{remarkES}
%(one may also include short wavelength modes $< \ell$)
\bea
\label{deltaH}
\delta H = -E_S \sum_{\cal B} S^+_{\cal B} S^-_{\cal B},
\eea
where $E_S \sim \frac{g^2 m}{\hbar^2}= \frac{\hbar}{\tau}$ with a numerical coefficient of $\mathcal{O}(1)$.
 On the superradiance line, $S_{\cal B}^z=0$, which will be our main concern from now on, Eq.~(\ref{deltaH}) reduces to $\delta H = -E_S \sum_{\cal B} \vec{S}_{\cal B}^2$, giving the energy difference between the largest and the next largest spin states, $E_S [S(S+1) - (S-1)S]=E_S N $. This justifies keeping only the large-$S$ manifold for low temperatures $T \ll E_S N$.
 %As shown in ~\cite{appendix}, this is not the case at the critical temperature, whose calculation must probe the full Hilbert space.

\emph{Dicke superradiance versus decoherence:} We finally include the coupling to the local vibrations in Eq.~(\ref{eq:Hmodel}). In general, a strong coupling to such a baths can destroy the quantum coherence needed for the Dicke effect and for superfluidity. To estimate the effects of decoherence, we focus on the region near the superradiance line where collective effects are most pronounced.

It is convenient to define $\Sigma(E)=\sum_l C_l^2/(E  - E_l+ i 0^+)$. For a single TLS, the decoherence rate  $\Gamma =- \frac{2 }{\hbar} {\rm{Im}} \Sigma(\Delta E)$.
 %is equivalent to the rate of relaxation between states $\sigma^x_i = \pm 1$ with the energy difference $\Delta E$ (In the superradiant phase $\Delta E = 2 g {\rm{Re}}\langle f^\dagger \rangle$).
 As an alternative to the superradiant phase studied above, we consider a state where all local oscillators deform into their new ground state at $|\langle \sigma^z_i \rangle|=1 $, and where $\langle f^\dagger \rangle=0$. The energy gain is $-\sum_l \frac{C_l^2}{E_l}=  {\rm{Re}} \Sigma(0)$ per TLS. The corresponding 2D energy density $E_{\rm{v}}=  {\rm{Re}} \Sigma(0) \rho_{TLS}d$, grows only linearly in $N = \rho_{TLS}dl^2$, as opposed to the superradiance phase with $E_{GS} \propto - N^2$, see Eq.~(\ref{eq:EGS}). The superradiance phase is the ground state if $E_{GS} \ll E_{\rm{v}}$. This gives, setting $\mu \sim \Omega$ in Eq.~(\ref{eq:EGS}),  $\Omega \gg |\rm{Re} \Sigma(0)|$.
 For a rough estimate, we assume that ${\rm{Im}} \Sigma(\Delta E)$ and $\rm{Re} \Sigma(0)$ are of the same order.
 %Specifically, this occurs within the simplifying assumption of constant $C_l$ and constant density of states $\sum_l \delta(E-E_l)$.
 In that case the above criterion becomes $\Omega \gg \hbar \Gamma$.

A more restrictive criterion should ensure that higher energy dark spin states with lower value of $\vec{S}_\mathcal{B}^2$ are weakly excited by $H_{\rm{vib}}$. Starting from the large-$S$ ground state $| S \rangle$, the Hamiltonian $H_{\rm{vib}}$ causes transitions into lower spin states $\sigma_i^z |S \rangle$ whith $\vec{S}_\mathcal{B}^2 = (S-1)S$. From Eq.~(\ref{deltaH}) all these states are separated by approximately the same energy gap $E_g=E_S N$ from the ground state. Neglecting the vibrational energy of $\mathcal{O}(1)$ in $\Sigma(E_g)$ the corrections to the ground state due to higher energy states are small if $\sum_l |C_l|^2 \ll E_S^2 \rho_{TLS}d$, which will again be satisfied for sufficiently large $N$.

Estimating $\Omega$ from the typical parameters in Eq.~(\ref{eq:numbers}), we obtain $\Omega \approx 0.028 \hbar \omega_{c}$,
%$c_{\rm eff} = 0.11 c$,
while the decoherence rate satisfies $\Gamma \approx \omega_{c}$. Thus $\hbar \Gamma > \Omega$, implying that Dicke physics is subdominant in Ref.~\cite{Klaers}. However, while decoherence for single TLSs is enormous, $\Gamma \tau = 10^5$, Dicke physics reduces this large number by few orders of magnitude. We expect that upon increasing the dye concentration or the distance $d$, Dicke physics can become dominant and allow one to observe the rich pattern predicted and discussed here.

%The decoherence Hamiltonian can be now projected into the large spin manifold, by using $\sigma^z_i=\frac{2}{N} S_z $. We can see that it gives vanishing contribution on the superradiance line. The operator $S_z$ causes small angle rotatins of the large spin in the $xy$ plane. In the basis $| m,S \rangle$ where $m$ is the eigenvalue of $\vec{S} \cdot \hat{n}$, decoherence gives transition rates between $| m=S,S \rangle \to | m=S-1,S \rangle$. Using the Fermi's golden rule this rate is given also by $\Gamma$. Thus on time scale $1/\Gamma$ there are jumps of the spin by angle $2\pi/N$. This random walk implies that the time for full flip to $-\hat{n}$ is given $N^2/\Gamma$, as compared to $1/\Gamma$ for a single TLS.

\paragraph{Acknowledgments:} We acknowledge discussions with S. Bar-Ad, N. Davidson, S. Diehl, B. Fischer, K. A. Kikoin, and M. Weitz.
The work was supported by the People Programme (Marie Curie Actions) REA-618188 (ES), research group 960 of the DFG (AR), and the Israel Science Foundation (ES and VF).

\section{Supplementary Material}

In this supplement we provide details for the calculation of the critical temperature and superfluid velocity, as well as the effective theory at the quantum phase transitions and the influence of disorder.

\section{Critical Temperature}
\vspace{-4mm}
We have demonstrated that the microcavity filled with dye molecules can be viewed as a collection of correlated regions  of area $\ell^2$, drawn as boxes in Fig.~1 in the main text, containing $N$ TLSs each. In this section, we will use this picture to estimate the critical temperature $T_c$ of the SF phase. Since correlations are limited to length $\sim \ell$ characterizing each box, $T_c$ is the same for all the boxes, and can be calculated for a single box. Since the interaction between TLSs $\propto e^{-|r|/\ell}$ is substantial between all pairs in the box, for the sake of an order of magnitude estimate of $T_c$ we will place all $N$ TLSs at the center of the box. Under this simplification the model for each box is just the multimode Dicke model studied in Ref.~\cite{Hepp}. Adopting their notation, the Hamiltonian reads
\bea
H_\mathcal{B} &=& \sum_{k} \nu_k a_k^\dagger a_k\nonumber \\
&+& \sum_{i=1}^N \left(\frac{\omega}{2} \sigma_i^z+\frac{1}{2 \sqrt{N}}  (\sum_k \gamma_k a_k \sigma_i^+ + h.c. )\right).
\eea
We define $-\mu = {\rm{min}}_k \{\nu_k\}$. Relating to our model, $\nu_k =\frac{\hbar^2 k^2}{2 m} -\mu$ and $\omega=-(\Delta +\mu)$. Wang and Hioe obtained an exact result for $T_c$, given by~\cite{Hepp}
\bea
\label{eq:WH1}
\tanh \left( \frac{\mu+\Delta}{2 T_c} \right) = \frac{- \mu (\mu+\Delta)}{\gamma^2},~~(|\mu(\mu+\Delta)| < \gamma^2)
\eea
and $T_c=0$ otherwise. Here $\gamma^2 = \sum_k \frac{{\rm{min}}_k \{\nu_k\}}{\nu_k}\gamma_k^2$.

We may also relate $\gamma_k$ to our $g$ via $\frac{\gamma_k}{2 \sqrt{N}} = \frac{g}{\ell}$. We used $f(r) = \frac{1}{\ell} \sum_k e^{i \vec{k} \vec{r}}a_k$. Thus
\bea
\label{eq:log}
\gamma^2 = \frac{4 N g^2}{\ell^2} \sum_k \frac{-\mu}{\frac{\hbar^2 k^2}{2m}-\mu} = \Omega^2 \frac{1}{\pi} \frac{-\mu}{\frac{\hbar^2}{2 m \ell^2}} \log \frac{\hbar \omega_c}{-\mu},
\eea
with $k=(k_x,k_y)$ and $k_x$ and $k_y$ are quantized in units of $\frac{2 \pi}{\ell}$.
For large positive $\Delta$, $\Delta \gg \Omega$  we may use Eq.~4 in the main text, $\frac{\hbar^2 }{2 m \ell^2} = \Delta$. In this limit the SF region is confined near the superradiant line $-\mu = \Delta$. Thus, up to a prefactor of order $1$ which includes the logarithmic factor in Eq.~(\ref{eq:log}), we simply have $\gamma \sim \Omega$, so from Eq.~(\ref{eq:WH1}) the maximal value of $T_c$ on the superradiance line is
\bea
\label{eq:Tc}
T_c \sim \frac{\Omega^2}{\Delta}, ~~~(\mu+\Delta=0,~~~\Delta \gg \Omega).
\eea
This result also gives the order of magnitude $T_c \sim \Omega$ of the transition temperature in the regime where $\Delta \approx \Omega$.

We notice that a simple mean field analysis of our model Eq. (1) in the main text gives directly Eq.~(\ref{eq:Tc}):
The mean field value for $\langle f \rangle=\langle f^\dagger \rangle$ gives, for $-\mu=\Delta$, $-\Delta \langle f \rangle=\rho_{\rm{TLS}} d g \langle \frac{\sigma^x}{2} \rangle$, with $\langle \frac{\sigma^x}{2} \rangle$ being the expectation value of a spin-1/2 in a magnetic field given by
$B=g \langle f \rangle$, therefore $\langle \frac{\sigma^x}{2} \rangle=-\frac{1}{2} \tanh \left( g \langle f \rangle/(2 T_c) \right)$. Linearizing in $\langle f \rangle$ gives the desired result. This mean field calculation confirms that the critical temperature of each box (in the limit $N \to \infty$) determines the critical temperature of the full system. We emphasize that in this finite temperature mean field calculation, in contrast to the $T=0$ case that will be described in the next section, it is important to retain all $2^N$ spin states apart from the space of $N
+1$ states of a single large spin. While they have higher energy they contribute to the entropy.

In the 2D system there is no ordering and instead there is a Kosterlitz-Thouless transition, but we expect that its critical temperature will be of similar size as the mean field critical temperature.
\vspace{-5mm}

\section{$T=0$ Mean Field and Sound Velocity}
\vspace{-2mm}
We return to the zero temperature case. Due to the energetic preference for large spin formation, see Eq.~(15) in the main text, we will treat the TLSs as large spins forming on typical regions of size $\ell$, $S=N/2=\ell^2 d \rho_{\rm{TLS}}/2$.
Interestingly, the value of $\ell$ will not influence any of the results in this section. It will, however, set the typical healing length over which sound waves form. The velocity of these collective excitations is calculated in this section.

Parametrizing the photon field as $f(r,t) e^{i \alpha(r,t)}$ (with $f>0$) and the Dicke spin as $S^+(r,t) = S \sin(\theta(r,t))e^{i \phi(r,t)}$, which are assumed to vary slowly on the scale $\ell$, the Lagrangian density $\mathcal{L}[f(r,t),\alpha(r,t),\theta(r,t),\phi(r,t)] $ becomes
\bea
\label{eq:MFl}
\mathcal{L}&=&- \hbar  f^2 \dot{\alpha} + \mu f^2 - \frac{\hbar^2 f^2}{2m} (\nabla_r \alpha)^2 - \frac{\hbar^2}{2m}(\nabla_r f)^2
 \\&-&\frac{S}{\ell^2}  \left(\hbar  \cos (\theta) \dot{\phi} -(\mu+ \Delta)\cos(\theta)+2 g \sin(\theta) f  \cos (\phi + \alpha) \right) .\nonumber
\eea
Note that this expression does not depend on $\ell$ as only the combination $\frac{S}{\ell^2} = \frac{d \rho_{\rm{TLS}}}{2}$ enters. The time-derivative term $\propto \cos(\theta) \dot{\phi}$ is the Berry phase describing the dynamics of a spin. The uniform ground state is obtained by minimizing the energy. This gives no restriction on $\phi-\alpha$, which is the massless Goldstone mode, and
\bea
\phi+\alpha&=& \pi,~~~(g>0) \nonumber \\
\mu f &=& \frac{g S}{\ell^2} \sin(\theta),\nonumber \\
\Delta+\mu &=& -2 g f \cot(\theta).
\eea
The latter two equations can be solved for $f=f_0$ and $\theta = \theta_0$. This gives
\be
\cos(\theta_0) =- \frac{\mu (\Delta + \mu)}{g^2 (2 S/\ell^2)}=-\frac{\mu (\Delta + \mu)}{\Omega^2},
\ee
and $f$ is given in Eq.~(12) in the main text. Deviations from $f=f_0$ and $\theta = \theta_0$ are massive fluctuations, in terms of which the energy density can be expanded to quadratic order,
\bea
E^{(2)} = E_{GS}+\frac{1}{2} \left(
\begin{array}{cc}
\delta f & \delta \cos \theta \\
\end{array}
\right) \hat{E}'' \left(
\begin{array}{c}
 \delta f \\
 \delta \cos \theta \\
 \end{array}
 \right) \nonumber ,
\eea
with
\bea
\hat{E}''&=&\left(
            \begin{array}{cc}
              \frac{\partial^2 E}{\partial f ^2} & \frac{\partial^2 E}{\partial f \partial \cos \theta} \\
              \frac{\partial^2 E}{\partial f \partial \cos \theta} & \frac{\partial^2 E}{\partial \cos \theta ^2} \\
            \end{array}
          \right)_{f=f_0,\theta = \theta_0} \nonumber \\
           &=& \left(
                                                \begin{array}{cc}
                                                  -2 \mu & \rho^{2D} g \cot \theta_0 \\
                                                  \rho^{2D} g \cot \theta_0 & \rho^{2D} g f_0/\sin^2 \theta_0 \\
                                                \end{array}
                                              \right)
,
\eea and the ground state energy density $E_{GS}$ is given in Eq.~(13) in the main text. We defined the 2D density of TLSs, $\rho^{2D}=\rho_{\rm{TLS}} d=
2 S / \ell^2$.
\vspace{-2mm}

To obtain the dynamics of the low energy Goldstone mode, we integrate out the massive fluctuations $\delta f$ and $\delta \cos \theta$, whose linear coupling to the time derivatives of $\alpha$ and $\phi$ in Eq.~(\ref{eq:MFl}) may be expressed as $\delta \mathcal{L}=-(\delta f~ \delta \cos \theta)\hat{M} (\dot{\alpha}~\dot{\phi})^T$ where $\hat{M}=\hbar \left(
                       \begin{array}{cc}
                         2  f_0 & 0 \\
                         0 & \rho^{2D}/2  \\
                       \end{array}
                     \right) $. Performing the integration gives the effective action
\bea
 e^{i\int d^2 r dt \mathcal{L}_{\rm{eff}}}&\propto & \int \mathcal{D} \delta f(r,t) \mathcal{D} \delta \cos \theta(r,t) e^{i\int d^2 r dt \mathcal{L}^{(2)}} ,\nonumber \\
\mathcal{L}_{\rm{eff}} &=& \frac{1}{2} \left(
                                       \begin{array}{cc}
                                         \dot{\alpha} & \dot{\phi}
                                          \\
                                       \end{array}
                                     \right)\hat{M}(\hat{E}'')^{-1}\hat{M} \left(
\begin{array}{c}
\dot{\alpha}  \\
\dot{\phi} \\
\end{array}
\right)  \\
&-& \frac{\hbar^2 f_0^2}{2m} (\nabla_r \alpha)^2 - \frac{2gS}{\ell^2} \sin(\theta_0) f_0 \cos(\phi+\alpha).\nonumber
\eea
Two phase fields $\alpha$ and $\phi$ are governed by this effective Lagrangian, among which there is one Goldstone mode $\phi_-=\phi-\alpha$ and one massive mode $\phi_+=\phi+\alpha$. In terms of $\phi_{\pm}(r,t)$ the Lagrangian reads
\bea
\label{Lphipm}
\mathcal{L}_{\rm{eff}} &=& -\frac{m_+}{2}  \phi_+^2+ \frac{1}{2} \left(
                                                    \begin{array}{cc}
                                                      \dot{\phi}_+ & \dot{\phi}_- \\
                                                    \end{array}
                                                  \right) \hat{A}  \left(
                                                                     \begin{array}{c}
                                                                       \dot{\phi}_+  \\
                                                                       \dot{\phi}_- \\
                                                                     \end{array}
                                                                   \right)\nonumber \\
                                                                   &-&  \frac{1}{2} \left(
                                                    \begin{array}{cc}
                                                      \nabla_r \phi_+ & \nabla_r \phi_- \\
                                                    \end{array}
                                                  \right) \hat{B}  \left(
                                                                     \begin{array}{c}
                                                                       \nabla_r \phi_+  \\
                                                                       \nabla_r \phi_- \\
                                                                     \end{array}
                                                                   \right),
\eea
with mass $m_+=\frac{2g  f_0 S \sin \theta_0 }{\ell^2}$, $\hat{A}=\frac{1}{4} \left(
                       \begin{array}{cc}
                         1 & 1 \\
                         -1 & 1 \\
                       \end{array}
                     \right)  \hat{M}(\hat{E}'')^{-1}\hat{M}\left(
                       \begin{array}{cc}
                         1 & -1 \\
                         1 & 1 \\
                       \end{array}
                     \right)
$ and $\hat{B} = \frac{\hbar^2 f_0^2}{4 m} \left(
                       \begin{array}{cc}
                         1 & -1 \\
                         -1 & 1 \\
                       \end{array}
                     \right) $.
Integrating out the gapped mode $\phi_+$ generates only fourth order derivative terms for $\phi_-$. Thus up to second order derivative terms of the gapless mode the Lagrangian is written as $\mathcal{L}_{\rm{eff}} = \frac{A_{22}}{2} \dot{\phi}_-^2 - \frac{B_{22}}{2} (\nabla_r \phi_-)^2 $, which describes a wave with velocity
\be
\label{eq:sBA}
s = \sqrt{B_{22} / A_{22}}.
 \ee
 Working out the algebra one obtains Eq.~(14) in the main text.

 Any sound wave requires interactions. Without the TLSs, a  sound mode made of photons can not be achieved in our model where photons are noninteracting. An effective photon-photon interaction, responsible for this sound mode, stems from the TLSs and will be calculated in the next section.

The velocity in Eq.~(\ref{eq:sBA}) exceeds the speed of light $c$ when the coupling $g$ is large enough. However, we will see now that this is not the case anymore if one starts from the correct theory of massive photons.

\subsubsection{Effective action from Maxwell's equations}

We demonstrate now that beyond the leading expansion in large $\omega_c$, which is typically considered, one obtains a second order time derivative $f^* \partial_t^2 f$ in the action of the massive photon. Consider the equation
\bea
[\partial_t^2 -c^2 (\nabla_z^2 + \nabla_\textbf{r}^2)]E(r,z,t)=0,
\eea
for the electric field in a linear medium with the speed of light $c$. It is obtained from the Maxwell's equation by excluding the magnetic component. We apply the ansatz
\bea
E(r,z,t) \propto f(r,t) e^{-i \omega_c t}\sin(n_z \pi z/d)+h.c..
\eea
It is generally assumed that it is sufficient to consider one polarization, determined by an external laser pump. The equation of motion for $f$ is
\bea
\label{eq:f}
{\rm{Re}} \left[ \partial_t^2 -2 i \omega_c \partial_t   -c^2 \nabla_\textbf{r}^2 \right]f=0.
\eea
Taking $f$ to have dimensions of inverse length our system is described by the Lagrangian
\bea
\label{eq:maxwell}
L=
\int d^2 r  f^*(r)  \left( i \hbar \partial_t-\frac{\hbar}{2 m c^2} \partial_t^2 + \frac{\hbar }{2 m} \nabla_\textbf{r}^2 \right) f(r),
\eea
where we used $\omega_c=mc^2$.

Typically~\cite{Carusotto} the second term $\propto \partial_t^2$ is neglected for frequencies $\omega \ll \omega_c$. However, returning to the calculation of the SF velocity, this term gives an additional second order time derivative $\delta L =f^2 \frac{\hbar}{2 mc^2} {\dot{\alpha}}^2$ in the effective Lagrangian Eq.~(\ref{eq:MFl}). Eventually, it gives  a minimal value for $A_{22}$ in Eq.~(\ref{Lphipm}) which becomes obtains an extra contribution $A_{22} \to A_{22} + \frac{\hbar^2 f_0^2}{4 mc^2}$. As a result, the sound velocity becomes
\bea
s = \sqrt{\frac{\frac{\hbar^2 f_0^2}{4 m}}{\frac{\hbar^2 f_0^2}{4 mc^2}+A_{22}}} <  c,
\eea
which never exceeds the speed of light.

\vspace{-2mm}
\section{Critical Theory}
The general critical theory near the transition lines in the phase diagram, Fig. 3 in the main text, is described by the Lagrangian density
\bea
\label{LN=0b}
\mathcal{L} = \psi^*  [c_1 i \hbar \partial_t+ c_2  \partial_t^2+\frac{\hbar^2}{ 2 m^*} \nabla_r^2+\delta \mu ]\psi -\frac{u}{2} |\psi|^4.
\eea
The transition lines $\mu = \mu_0$ or $\mu=\mu_\pm$ correspond to $\delta \mu=0$. The quadratic part of the theory follows from Eqs.~(7) and (9); the dispersion relation of two modes is obtained from $\det \mathcal{L}_{0,1}(\omega (k), k)=0$, and the critical theory is obtained by retaining only the gapless mode for which $\omega(k=0)=0$ at $\delta \mu =0$. Generic phase transitions in Fig.~3 in the main text have quadratic dispersion $\omega(k) \sim k^2$, corresponding to dynamical critical exponent $z=2$. This behavior applies to the entire transition line $\mu = \mu_0$.
 %(and interactions are marginally irrelevant in $d+z=4$), and then $c_2$ is irrelevant.
A characteristic feature of the Mott transition~\cite{Sachdev} is, however, that there is always one point with $c_1=0$, and hence $z=1$ and the critical mode has linear dispersion $\omega(k)=c_{\rm eff} k$ with an effective speed of light $c_{\rm eff}=\sqrt{\frac{\hbar^2}{ 2 m^* c_2}}$. This Lorentz invariant quantum critical point was identified at $\Delta= 2 \Omega$, $\mu=-\Omega$ (red point in Fig.~3 in the main text), where one can obtain from $\det \mathcal{L}_{1}(\omega (k), k)=0$ that $c_{\rm eff}=\left( \frac{\Omega}{2m}\right)^{1/2}$.

In this section we will focus on the transition between the vacuum phase and SF phase ($\mu=\mu_0$ in Fig.~3 in the main text) and calculate the parameters of the critical theory Eq.~(\ref{LN=0b}). Later, we will study the influence of disorder at this transition.

The dispersion relation $\omega(k)$ is determined from
 \bea
 \det \left(
        \begin{array}{cc}
          \omega(k)-\frac{\hbar^2 k^2}{2m}+\mu_0+\delta \mu & -\Omega \\
          -\Omega & \omega(k)+\mu_0+\delta \mu+\Delta \\
        \end{array}
      \right)=0. \nonumber
 \eea
This equation gives two eigenmodes being linear combinations of the photon- and Dicke fields: (i)  The critical field is the linear combination $\psi(r,t) = \cos(\eta) f(r,t)+\sin (\eta) D(r,t)$ which has zero energy at criticality, $\omega=k=0$, $\mu=\mu_0$,
\be
\left(
  \begin{array}{cc}
    \mu_0 & -\Omega \\
    -\Omega & \mu_0 + \Delta \\
  \end{array}
\right) \left(
          \begin{array}{c}
            \cos(\eta) \\
            \sin(\eta) \\
          \end{array}
        \right),
\ee
giving $\tan (\eta) = \mu_0/\Omega$. (ii) The orthogonal combination  $\phi = -\sin (\eta) f+\cos(\eta) D$ is the non-critical mode which has an energy gap at the transition. The chemical potential couples to the total conserved density $D^* D+f^* f = \psi^* \psi+ \phi^* \phi$. Near the quantum phase transition the non-critical mode $\phi$ will be neglected. This way we find $c_1=1$, $c_2=0$ (the contribution from Eq.~(\ref{eq:maxwell}) to $c_2$ can be neglected in a $z=1$ QPT), $\delta \mu = \mu-\mu_0$, and
\be
\frac{1}{m^*}=\frac{1}{\hbar^2} \partial_k^2 \omega(k) =\frac{1-\frac{\Delta}{\sqrt{\Delta^2+4 \Omega^2}}}{2m}.
\ee
\emph{Interaction parameter:} While photons are noninteracting in our model, a finite interaction $u$ in Eq.~(\ref{LN=0b}) exists due to the presence of the TLSs. We will use now our $T=0$ mean field results to obtain $u$.

The interaction $u$ can be obtained in various equivalent ways.
The interaction parameter $u$ determines the ratio between changes in the chemical potential and SF density; from Eq.~(\ref{LN=0b}) we have $u = \lim_{\delta \mu \to 0} \left( \frac{\langle \psi^* \psi \rangle }{\delta \mu}\right)^{-1}$. (Another equivalent way to obtain $u$ is by using the relation $m^* s^2 = u \psi^* \psi$ which gives the same result). We may use our knowledge of $\langle f^* f \rangle$ and $\sqrt{{S^x}^2+{S^y}^2}$ from the mean field solution to evaluate $\langle \psi^* \psi \rangle$, via $\langle \psi^* \psi \rangle  = (S+S^z)/\ell^2+  \langle f^* f \rangle   = S/\ell^2(1+ \cos \theta_0)+  f_0^2$, where $f_0$ and $\theta_0$ were evaluated in the previous section. We consider the dimensionless coupling constant $\tilde{u}=\frac{2m^* u }{\hbar^2}$, which is equal to $(\langle \psi^* \psi \rangle l_h^2)^{-1}$ and
\be
\label{eq:lh}
l_h =\frac{\hbar}{\sqrt{2m^* \delta \mu}},
\ee
is the healing length. Along the transition to the vacuum phase we find
\bea
\label{eq:tildeu}
\tilde{u} = \frac{m(\Delta+\sqrt{\Delta^2+4 \Omega^2})^2}{\hbar^2 d \rho_{\rm TLS} \sqrt{\Delta^2+4 \Omega^2}}.
\eea
In Fig.~(\ref{fg:1}) we plot $\tilde{u}$ as function of $\Delta$, using the parameters in Eq.~(2) in the main text and see that very small interaction parameter is obtained everywhere. We emphasize, however, that the manifestation of superfluidity, e.g., formation of sound waves, requires distances which exceed the healing length. Setting $\delta \mu \sim \Omega \sim \Delta$, and multiplying and dividing Eq.~(\ref{eq:lh}) by $\ell^2$ using Eq.~(4) from the main text, we see that
\be
l_h \sim \ell.
 \ee
 Physically this length is the distance at which correlation is built. The fact that $\tilde{u}$ is small simply reflects the fact that density $\psi^* \psi = \delta \mu/u$ responds strongly to changes in the chemical potential. This is consistent with the dimensionality mismatch between the 2D photons and 3D molecules, since, for typical changes of $\delta \mu$ being of the order of $\Omega$, we will obtain a SF density consisting of a fraction of the TLSs, $\rho \sim \rho_{\rm{TLS}} d$, giving again $\tilde{u} \sim \frac{2m}{\hbar^2}\frac{\Omega}{\rho_{\rm{TLS} }d} \sim \frac{1}{\rho_{\rm{TLS}} d \ell^2} \sim 8 \cdot 10^{-6}$ observed in Fig.~4 for $\Delta \to 0$.

\begin{figure}[h]
\centering
\includegraphics*[width=8cm]{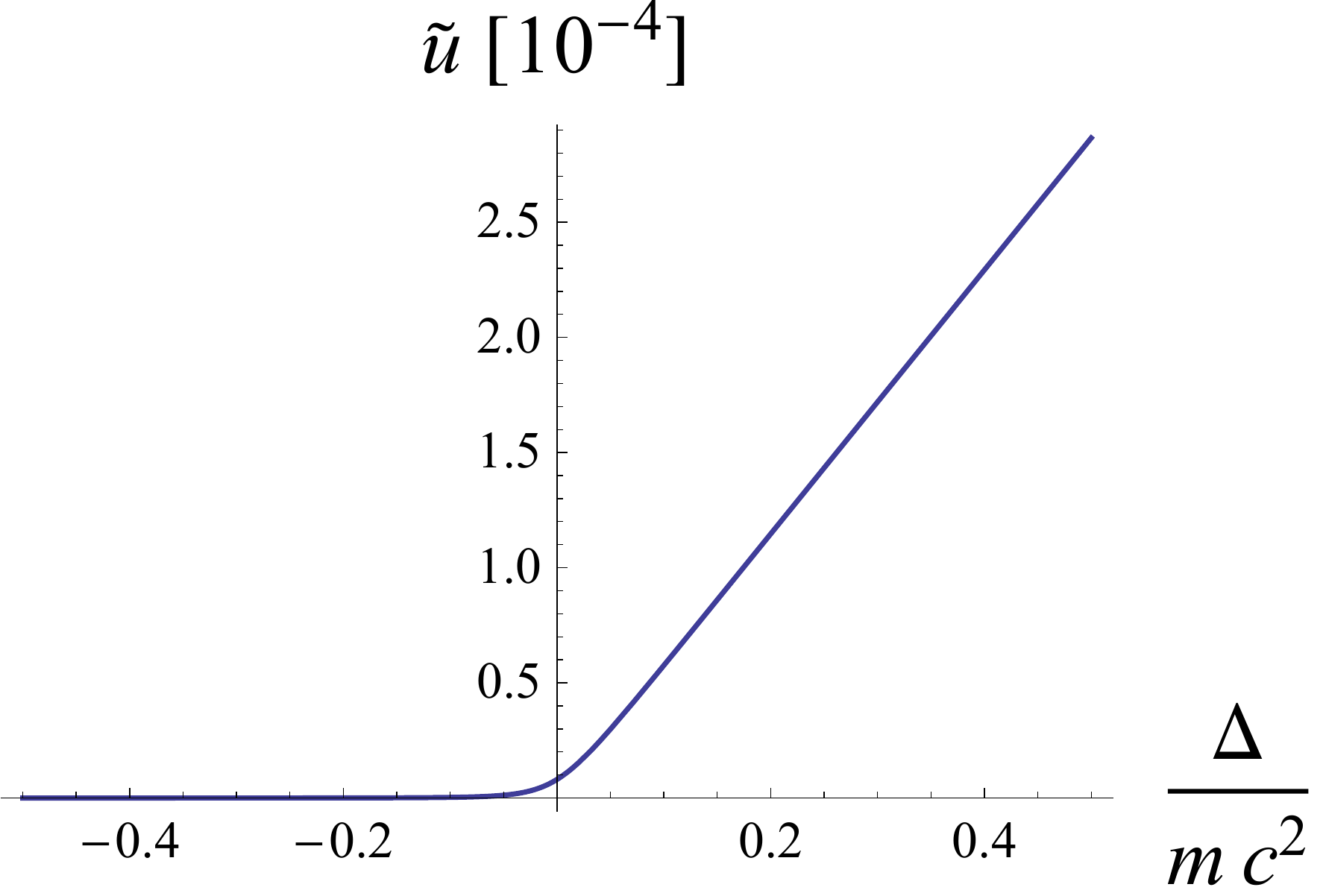}
%\vspace{-1cm}
\caption{\label{fg:1} Dimensionless interaction $\tilde{u}$, Eq.~(\ref{eq:tildeu}), as function of $\Delta$ along the transition line $\mu = \mu_0$. We used $\Omega = 0.028 m c^2$.
}
\end{figure}
In the photon condensation limit, where $\langle \psi^* \psi \rangle \to \langle  f^* f \rangle$, the interaction is given by $u = \frac{2  {g}^4}{ \Delta^3} d \rho_{\rm TLS}$, that corresponds to a local contribution of $\frac{2  {u}^4}{ \Delta^3}$ from each TLS, and is thus proportional to the 2D density of molecules $\rho_{\rm TLS} d$.

\emph{Influence of disorder near the transition:}
The random arrangement of TLSs gives a source of disorder which we have ignored in our coarse grained description. A simple order of magnitude estimate does, however, show that disorder can safely be neglected, as we describe now.

The number of TLSs within one box in Fig. 1 in the main text fluctuates by $ \pm \sqrt{N}$ around its mean value $N$. The energy scale $\Omega$ scales as $\Omega \propto \sqrt{N}$; hence fluctuations in $\Omega^2$ go like $\Omega^2/\sqrt{N}$. Assume now that $\delta \mu$ is tuned near the critical point. $\delta \mu$ depends linearly on $\mu_0(\Omega)$. Expanding this function at large $\Delta/\Omega$ we have $\mu_0 \sim -\Delta - \frac{\Omega^2}{\Delta}$. Thus we obtain fluctuations in the tuning parameter $\delta \mu$ given by $\frac{\Omega^2}{\Delta \sqrt{N}}$. Now, $N$ is related to length $L$ via $N(L) = d \rho_{\rm{TLS}} L^2$. Thus, fluctuations of $\delta \mu^{(2)} =\sqrt{  \langle  \delta \mu^2 \rangle }$ are generated at length
\be
L = \frac{\Omega^2}{\Delta \sqrt{\rho_{TLS}d}}\frac{1}{\delta \mu^{(2)}}.
 \ee
Comparing $L$ with the correlation length at distance $\delta \mu = \delta \mu^{(2)}$ from criticality,
\be
\xi =\frac{\hbar}{\sqrt{2m^*}} \frac{1}{\sqrt{\delta \mu^{(2)}}},
 \ee
 one obtains, from the Harris criterion, that the disorder can be neglected for $\frac{\Omega^4}{\Delta^3} \frac{2m^*}{\hbar^2 \rho_{\rm{TLS }}d} < |\mu-\mu_0|$. Multiplying and dividing by $\ell^2$, using $\Delta = \frac{\hbar^2}{2 m \ell^2}$, we have $\frac{\Omega^4}{\Delta^3} \frac{1}{N} < |\mu-\mu_c|$.  Due to the factor $1/N$, disorder can be ignored for realistic parameters, except for a very narrow region near the quantum phase transition line.


\begin{thebibliography}{21}
\expandafter\ifx\csname natexlab\endcsname\relax\def\natexlab#1{#1}\fi
\expandafter\ifx\csname bibnamefont\endcsname\relax
  \def\bibnamefont#1{#1}\fi
\expandafter\ifx\csname bibfnamefont\endcsname\relax
  \def\bibfnamefont#1{#1}\fi
\expandafter\ifx\csname citenamefont\endcsname\relax
  \def\citenamefont#1{#1}\fi
\expandafter\ifx\csname url\endcsname\relax
  \def\url#1{\texttt{#1}}\fi
\expandafter\ifx\csname urlprefix\endcsname\relax\def\urlprefix{URL }\fi
\providecommand{\bibinfo}[2]{#2}
\providecommand{\eprint}[2][]{\url{#2}}

\bibitem[{Dic()}]{Dicke54}
\bibinfo{note}{R. H. Dicke, Phys. Rev. \textbf{93}, 99 (1954)}.

\bibitem[{Hep()}]{Hepp}
\bibinfo{note}{K. Hepp and E. Lieb, Ann. Phys \textbf{76}, 360 (1973); Y.
  K.Wang, and F. T. Hioes, Phys. Rev. A \textbf{7}, 831(1973)}.

\bibitem[{ess()}]{esslinger}
\bibinfo{note}{K. Baumann, C. Guerlin, F. Brennecke, and T. Esslinger, Nature
  (London) \textbf{464}, 1301 (2010)}.

\bibitem[{Kla({\natexlab{a}})}]{Klaers}
\bibinfo{note}{J. Klaers, J. Schmitt, F. Vewinger and M. Weitz, Nature
  \textbf{468}, 545 (2010)}.

\bibitem[{Car()}]{Carusotto}
\bibinfo{note}{I. Carusotto, C. Ciuti, Rev. Mod. Phys. \textbf{85}, 299
  (2013)}.

\bibitem[{Kir()}]{Kirton13}
\bibinfo{note}{P. Kirton and J. Keeling, Phys. Rev. Lett. \textbf{111}, 100404
  (2013) .}

\bibitem[{Fis()}]{Fischer}
\bibinfo{note}{B. Fischer and R. Weill, Opt. Express \textbf{20}, 26704
  (2012).}

\bibitem[{Mar()}]{Martini}
\bibinfo{note}{E. De Angelis, F. De Martini, and P. Mataloni, J. Opt. B
  \textbf{2}, 149 (2000).}

\bibitem[{Gre()}]{Greentree06}
\bibinfo{note}{Greentree, A. D., C. Tahan, J. H. Cole, and L. C. L. Hollenberg,
  Nature Phys. \textbf{2}(12), 856 (2006)}.

\bibitem[{Dim()}]{Dimer07}
\bibinfo{note}{F. Dimer, B. Estienne, A. S. Parkins, and H. J. Carmichael,
  Phys. Rev. A \textbf{75}, 013804 (2007).}

\bibitem[{Dal()}]{DallaTorre13}
\bibinfo{note}{E. G. Dalla Torre, S. Diehl, M. D. Lukin, S. Sachdev and P.
  Strack, Phys. Rev. A \textbf{87}, 023831 (2013).}

\bibitem[{Buc()}]{Buchhold13}
\bibinfo{note}{M. Buchhold, P. Strack, S. Sachdev, and S. Diehl, Phys. Rev. A
  \textbf{87}, 063622 (2013).}

\bibitem[{Kla({\natexlab{b}})}]{Klaers1}
\bibinfo{note}{J. Klaers, J. Schmitt, T. Damm, F. Vewinger, and M. Weitz, Appl.
  Phys. B \textbf{105}, 17 (2011)}.

\bibitem[{Hal()}]{Haldane76}
\bibinfo{note}{F. D. M. Haldane and P. W. Anderson Phys. Rev. B \textbf{13}
  2553 (1976)}.

\bibitem[{Fle()}]{Fleurov76}
\bibinfo{note}{V. N. Fleurov and K. A. Kikoin, J. Phys. C: Solid State Phys.
  \textbf{9} 1673 (1976)}.

\bibitem[{Del()}]{Delta0}
\bibinfo{note}{At $\Delta=0$ the equation $\det [(E+\Delta) \delta_{ij} -g_i
  I_{ij}(E) g_j]=0$ gives an avoided crossing, and the length scale $\ell$
  saturates at $\ell \sim \frac{\hbar^2}{m g \sqrt{N}} \sim
  \frac{\hbar}{\sqrt{2 m \Omega}}$.}

\bibitem[{app()}]{appendix}
\bibinfo{note}{See supplementary material.}

\bibitem[{Sac()}]{Sachdev}
\bibinfo{note}{S. Sachdev, Quantum phase transitions, Cambridge University
  Press (2001)}.

\bibitem[{Hig()}]{Higgs}
\bibinfo{note}{M. Endres, T. Fukuhara, D. Pekker, M. Cheneau, P. Schauss, C.
  Gross, E. Demler, S. Kuhr and I. Bloch, Nature \textbf{487}, 454 (2012).}

\bibitem[{rem({\natexlab{a}})}]{remark}
\bibinfo{note}{Near the transitions assuming constant couplings merely enhances
  $g_{\rm{eff}}$ by a factor $\sqrt{2}$. We conjecture that this assumption
  captures the qualitative features also inside the SF region.}

\bibitem[{rem({\natexlab{b}})}]{remarkES}
\bibinfo{note}{Integrating out the photon band from scale $D$ up to the
  cut-off, as in Eq.~(3), one obtains $E_S= \frac{g^2 m}{2 \pi \hbar^2} \log
  \frac{\hbar \omega_{c}-\mu}{D-\mu}$.}

\end{thebibliography}
\end{document}